\documentstyle[sprocl]{article}

\def\be{\begin{equation}}
\def\ee{\end{equation}}
\def\bea{\begin{eqnarray}}
\def\eea{\end{eqnarray}}

\input epsf
\input colordvi  

\def\slashchar#1{\setbox0=\hbox{$#1$}           
   \dimen0=\wd0                                 
   \setbox1=\hbox{/} \dimen1=\wd1               
   \ifdim\dimen0>\dimen1                        
      \rlap{\hbox to \dimen0{\hfil/\hfil}}      
      #1                                        
   \else                                        
      \rlap{\hbox to \dimen1{\hfil$#1$\hfil}}   
      /                                         
   \fi}                                         %
 
\begin{document}

\title{THE STANDARD MODEL AND THE LATTICE}

\author{MICHAEL CREUTZ}

\address{Physics Department, Brookhaven National Laboratory,\\
Upton, NY 11973, USA\\email: creutz@bnl.gov}


\maketitle\abstracts{I discuss some of the difficulties with
formulating chiral symmetry on the lattice and review a recently
proposed scheme for a fully finite and exactly gauge invariant lattice
regularization of the standard model.}

\section{Introduction}
Lattice gauge theory, now in its third decade, has since its inception
been plagued by difficulties with fermions.  There are two apparently
distinct fermion problems.  First is the issue of algorithms, arising
since fermionic fields are anti-commuting variables.  Since the
exponentiated action is an operator in a Grassmann space, comparisons
with random numbers for a Monte Carlo program are problematical.
Several ways around this have been devised, mostly based on
integrating the fermions analytically as a determinant, but in my
opinion the approaches remain awkward.  Furthermore, when there is a
background baryon density, {\it i.e.} a chemical potential term in the
action, cancelations between terms of varying phase make the problem
essentially unsolved except on the tiniest lattices.  This is not a
unique problem to lattice gauge theory; studying doping in many
electron models has equivalent difficulties.

In this talk, however, I concentrate on the other fermion issue:
chiral symmetry and doubling.  Since the lattice is a first principles
approach to field theory, one could ask why care about the details of
chiral symmetry.  Just put the problem on the computer, predict
particle properties, and they should come out correctly if the
underlying dynamics is relevant.  While this may perhaps be a logical
point of view, it ignores a vast lore built up over the years in
particle physics.  In the context the strong interactions, the pion
and the rho mesons are made of the same quarks, the only difference
being whether the spins are anti-parallel or parallel.  Yet the pion,
at 140 MeV, weighs substantially less than the 770 MeV rho.  Chiral
symmetry is at the core of the conventional explanation.  Since the up
and down quarks are fairly light, we have an approximately conserved
axial vector current, and the pion is believed to be the remnant
Goldstone boson of a spontaneous breaking of this chiral symmetry.

Another motivation for studying chiral issues on the lattice arises
when considering the weak interactions.  Here we are immediately faced
with the experimental observation of parity violation, neutrinos are
left handed.  In the standard electroweak model, fundamental gauge
fields are coupled directly to chiral currents.  The corresponding
symmetries are gauged, {\it i.e.} they become local, and are crucial
to the basic structure of the theory.  Since the lattice is the one
truly non-perturbative regulator for defining a field theory, if one
cannot find a lattice regularization for the standard model, the
standard model itself may not be well defined.

Regulating divergences via a lattice is by no means a unique process.
However, Wilson's original formulation \cite{wilson74} has some rather
remarkable properties when applied to strong \Red{q}uark
\Green{c}onfining \Blue{d}ynamics, usually called QCD.  First, the
approach is indeed a regulator: it makes the theory fully finite.
Second, the cutoff is non-perturbative, unlike more conventional
approaches which begin by finding a formally divergent Feynman diagram
and then cutting it off.  But diagrams are the basis of perturbation
theory.  The advantage of the lattice is its imposition before any
expansion.  Third, and perhaps the most remarkable, the Wilson
approach accomplishes the above two feats while maintaining an exact
local gauge symmetry.  Besides its inherent elegance, this precludes
the need for any gauge variant counter-terms in the renormalization
procedure.  Since the theory is fully finite at the outset, the whole
issue of gauge fixing is circumvented.

Given these features, it is natural to ask if a similar scheme exists
for the full standard model, including the gauged chiral currents.
The answer to this is, I believe, unknown.  Nevertheless, I will lead
this talk towards a scheme that may provide all of the above features.
It involves some rather complex couplings, opening possible routes to
failure, but does appear to include the necessary features for such a
formulation.

\section{What is chiral symmetry?}

For pedagogy I digress briefly on what I mean by chiral symmetry.  The
issues here are deeply entwined with representations of the Lorentz
group -- zero mass particles are special.  In particular, the helicity
of a massless particle is invariant under Lorentz transformations.
Each helicity state forms a separate representation of the Lorentz
group; for spin one-half the Dirac field can be separated into right
and left handed parts, $\psi_R$ and $\psi_L$, formally independent.
Furthermore, because of the way the Dirac matrices appear, the
helicity of a fermion is naively preserved under gauge interactions.
When a polarized electron at high energy scatters off of an
electromagnetic field, its helicity is unchanged.

\begin{figure}
\epsfxsize .3\hsize
\centerline{\epsfbox{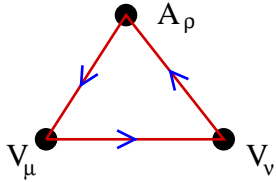}}
\caption{The triangle diagram cannot be regulated so both
vector and axial vector currents are conserved.}
\label{fig:triangle}
\end{figure}

This whole issue, however, is clouded by the so-called ``chiral
anomalies.''  In particular, the famous triangle diagram, sketched in
Fig.~\ref{fig:triangle}, coupling two vector and one axial vector
current is divergent, and no regularization can keep them both
conserved.  If either is coupled to a gauge field, such as the
electromagnetic current to the vector current, this diagram must be
regulated with that particular current being conserved.  Then the
other cannot be.  These anomalies are at the core of the lattice
problems.

The concept of chirality becomes even simpler in one spatial
dimension, where the direction of motion of a massless particle is
invariant under boosts.  Then the anomaly is easily understood via
simple band theory.\cite{anomalyflow}  A particle of non-zero mass $m$
and momentum $p$ has energy $E=\pm\sqrt{p^2+m^2}$.  Here I use a Dirac
sea description where the negative energy states are to be filled in
the normal vacuum.  Considering the positive and negative energy
states together, the spectrum of the system has a gap equal to twice
the particle mass.  In the vacuum the Fermi level is at zero energy,
exactly in the center of this gap.  In conventional band theory
language, the vacuum is an insulator.

\begin{figure}
\epsfxsize .6\hsize
\centerline {\epsfbox{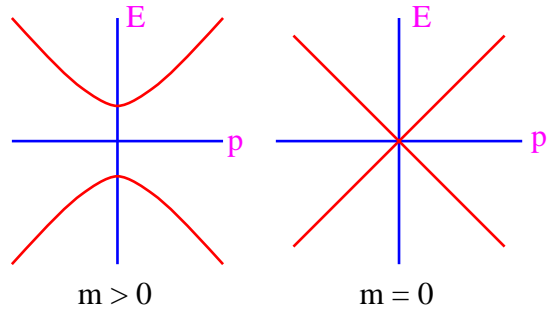}}
\caption{In one dimension the spectrum of massive particles has a gap,
and the vacuum can be regarded as an insulator.  The massless case, in
contrast, represents a conductor.  The anomaly manifests itself in the
ability to induce currents in a wire.}
\label{fig:conductor}
\end{figure}

In contrast, for massless particles where $E=\pm|p|$, the gap
vanishes.  The system becomes a conductor, as sketched in
Fig.~\ref{fig:conductor}.  Of course, conductors can carry currents,
and here the current is proportional to the number of right moving
particles minus the number of left movers.  If we consider gauge
fields, they can induce currents, a process under which the number of
right or left movers cannot be separately invariant.  This is the
anomaly, without which transformers would not work.

This induction of currents is not a conversion of particles directly
from left into right movers, but rather a sliding of levels in and out
of the Dirac sea.  The generalization of this discussion to three
spatial dimensions uses Landau levels in a magnetic field; the lowest
Landau level behaves exactly as the above one dimensional
case.\cite{anomalyflow}

One particularly important consequence for the standard model is that
baryon number is one of the anomalous charges.  Indeed,
't~Hooft\cite{thooft} pointed out a specific baryon-number-changing
mechanism through topologically non-trivial gauge configurations.  The
rate is highly suppressed due to a small tunneling factor and is far
too small to observe experimentally.  Nevertheless, the process is
there in principle, and any valid non-perturbative formulation of the
standard model must accommodate it.  If we have a fully finite and
exactly gauge invariant lattice theory, the dynamics must contain
terms which violate baryon number.  This point was emphasized some
time ago by Eichten and Preskill \cite{eichtenpreskill} and further by
Banks.\cite{banks}

Without baryon violating terms, something must fail.  In naive
approaches to lattice fermions the problem materializes via extra
particles, the so-called doublers, which cancel the anomalies.  For
the strong interactions alone, a vector-like theory, Wilson
\cite{wilson77} showed how to remove the doublers by adding a chirally
non-symmetric term.  This term formally vanishes in the continuum
limit, but serves to give the doublers masses of order the inverse
lattice spacing.  As chiral symmetry is explicitly broken, the chiral
limit of vanishing pion mass is only obtained with a fine tuning of
the quark mass, which is no longer ``protected'' with the bare and
physical quark masses no longer vanishing together.  This approach
works well for the strong interactions, but explicitly breaks a
chirally coupled gauge theory.  While perhaps tractable,\cite{rome} it
requires an infinite number of gauge variant counter-terms to restore
gauged chiral symmetries in the continuum limit.  It is these features
that drive us to search for a more elegant formulation.

To proceed I frame the discussion in terms of extra space-time
dimensions.  The idea of adding unobserved dimensions is an old one in
theoretical physics, going back to Kaluza and Klein,\cite{kaluzaklein}
and often is quite useful in unifying different interactions.  For my
purposes, it allows me to separate different parts of the problem, but
is probably only a theoretical crutch that can be removed at a later
stage.  Of course the extension of space-time to higher dimensions is
crucial to modern string theories.  Indeed, there are probably
unexploited analogies here, in particular chiral symmetry can become
quite natural when formulated on higher dimensional membranes.  Here I
use only the simplest extension, involving one extra dimension.

The use of an extra dimension in the context of anomalies also appears
in the area of effective chiral Lagrangians.  Here the famous
Wess-Zumino \cite{wesszumino} term is formulated in terms of an added
fifth dimension.  Anomalies in four dimensional currents are
interpreted as a flow into the fifth direction.  Indeed, the analogy
with the following discussion is striking, and recent arguments
\cite{mctytgat} suggest a close connection between the doubling
problem in lattice gauge theory and the problem of coupling gauge
fields to the Wess-Zumino term.

\begin{figure}
  \epsfxsize .5\hsize 
\centerline{\epsfbox{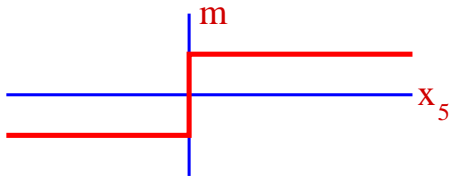}}
\caption{A step in a five dimensional fermion mass can give rise
to topological zero-energy fermion modes bound to a four dimensional 
interface.}
\label{fig:massstep}
\end{figure}

I start with an observation of Callan and Harvey,\cite{callanharvey}
building on Jackiw and Rebbi.\cite{jackiwrebbi} They argue that a five
dimensional massive fermion theory formulated with an interface where
the fermion mass changes sign, as sketched in Fig.~\ref{fig:massstep},
can give rise to a four dimensional theory of massless fermionic modes
bound to the interface.  The low energy states on the interface are
naturally chiral, and anomalous currents are elegantly described in
terms of a flow into the fifth dimension.

\begin{figure}
\epsfxsize .5\hsize
\centerline {\epsfbox{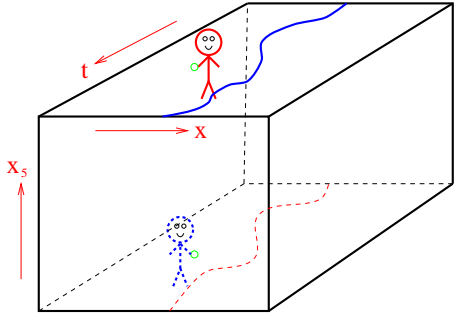}}
\caption{Regarding our four dimensional world as a surface in five
dimensions.}
\label{fig:kaplan}
\end{figure}

While the Callan and Harvey discussion is set in the continuum, Kaplan
\cite{kaplan} suggested carrying the formalism directly over to the
lattice.  His motivation, as mine here, is to provide a potential
scheme for chiral gauge theories on the lattice.  With the Wilson
formulation, the particle mass is controlled via the hopping
parameter, usually denoted $K$.  The massless situation is obtained at
a critical hopping, $K_c$, the numerical value of which depends on the
gauge coupling.  Thus, to set up an interface as used by Callan and
Harvey, one should consider a five dimensional theory with a hopping
parameter which depends on the extra fifth coordinate.  This
dependence should be constructed to generate a four dimensional
interface separating a region with $K>K_c$ from one with $K<K_c$.
Shamir \cite{shamir} observed a substantial simplification on the
$K<K_c$ side by putting $K=0$.  Then that region decouples, and the
picture reduces to a four dimensional surface of a five dimensional
crystal.  The physical picture is sketched in Fig.~\ref{fig:kaplan}.
For a Hamiltonian discussion, see Ref.~\cite{mcih}.  Indeed, surface
modes are not a particularly new concept; in 1939 Shockley
\cite{shockley} discussed their appearance in band models when the
interband coupling becomes strong.

The purpose of the fifth dimension is to address fermionic issues.
Since the bosonic sector of lattice gauge theory is already in good
shape, I require that the gauge fields not directly see the extra
coordinate.  In particular, any gauge field $A(x_\mu,x_5)=A(x_\mu)$ is
considered to independent of $x_5$.  Furthermore, the gauge field has
no fifth component, {\it i.e.} $A_5=0$.  A possibly helpful
analogy\cite{nn} regards $x_5$ as effectively a ``flavor'' index, with
hopping through the extra dimension representing a somewhat complex
``mixing.''

This approach gives a natural chiral theory on one wall of our system.
However, as Fig.~\ref{fig:kaplan} hints, for a finite five dimensional
system there are generally additional four dimensional surfaces.  One
extra surface, as in the figure, I refer to as an ``anti-wall.''
Indeed, on any finite system these topological defects occur in pairs.
This raises a difficulty with the above scheme for inserting the gauge
fields.  Since the latter do not know about the extra dimension, they
couple equally to modes on all walls.  On the anti-wall there are also
fermion zero modes, of the opposite chirality to those on the original
wall.  These are effectively ``mirror'' fermions; corresponding to a
left handed neutrino on the original wall, a right handed neutrino
appears on the anti-wall.  These mirrors cannot be neglected since they
couple with equal strength to the gauge fields.

Furman and Shamir \cite{furmanshamir} have argued that for vector-like
theories, such as the usual strong interactions, such a formulation
could be of considerable practical value.  In this case non-anomalous
chiral symmetries, responsible for the lightness of the pion, would be
natural in the limit of a large fifth dimension.  Indeed, the doubling
appearing with the anti-walls is in some sense the minimal required by
famous no-go theorems \cite{nielsenninomiya} Preliminary efforts with
this scheme have been promising.\cite{blumsoni}

Here, however, I am interested in chiral gauge theories such as the
standard model.  One might imagine eliminating the extra walls by
moving them off to infinity.  This lies at the heart of the closely
related ``overlap'' formalism of Ref.~\cite{nn} and provides a
non-perturbative definition for the chiral determinant.  However, how
anomaly cancelation works in this formulation is somewhat hidden, with
baryon non-conservation being relegated off to infinity.  Because of
this, in a recent paper \cite{us} we pursued an alternative scheme for
eliminating the doublers on the anti-wall.  We argued that a large
four-fermion coupling on the anti-wall could induce a mass gap of
order the lattice spacing for the spectrum on the ``bad'' wall.  The
form of the interaction has the same structure as the ``~`t Hooft''
vertices responsible for the baryon non-conservation discussed above
plus similar terms to break the anomalous strong axial $U(1)$
symmetry.  Physically, we give the mirror protons mass by mixing them
with the mirror positrons.  The primary danger is that the four
fermion interaction might spontaneously break one of the gauge
symmetries.  Such a breaking would naturally be at the scale of the
lattice spacing and would destroy the model.

Rather than describing the contents of that paper in more detail, I now
pursue an alternative but equivalent picture.  I am motivated by the
desire to understand how the no-go theorems are avoided; in particular
I discuss how all basic particles can be paired so every left handed
particle has a right handed counterpart.  This approach, also
discussed in my contribution to the Lattice '97
conference,\cite{lat97} involves a rather unusual reinterpretation of
the standard model.

\def \half {{\scriptstyle {1\over 2}}}

The standard model of elementary particle interactions is based on the
product of three gauge groups, $SU(3)\times SU(2) \times U(1)_{em}$.
Here the $SU(3)$ represents the strong interactions of quarks and
gluons, the $U(1)_{em}$ corresponds to electromagnetism, and the
$SU(2)$ gives rise to the weak interactions.  I ignore here the
technical details of electroweak mixing.  The full model is, of
course, parity violating, as necessary to describe observed helicities
in beta decay.  This violation is normally considered to lie in the
$SU(2)$ of the weak interactions, with both the $SU(3)$ and
$U(1)_{em}$ being parity conserving.  However, this is actually a
convention, adopted primarily because the weak interactions are small.
I argue below that a reassignment of degrees of freedom allows an
interpretation where the $SU(2)$ gauge interaction is vector-like.
Since the full model is parity violating, I must shift the parity
violation into the strong, electromagnetic, and Higgs interactions.

With a vector-like weak interaction, the chiral issues move to the
other gauge groups.  Requiring gauge invariance for the re-expressed
electromagnetism then clarifies the mechanism behind our above
mentioned proposal for a lattice regularization of the standard
model.\cite{us}

To see how this works, consider only the first generation, involving
four left handed doublets.  These correspond to the neutrino/electron
lepton pair plus three colors for the up/down quarks
\begin{equation}
\pmatrix{\nu \cr e^-\cr}_L,
\ \pmatrix{{u^r} \cr {d^r}\cr}_L,
\ \pmatrix{{u^g} \cr {d^g}\cr}_L,
\ \pmatrix{{u^b} \cr {d^b}\cr}_L
\end{equation}
Here the superscripts from the set $\{r,g,b\}$ represent the internal
$SU(3)$ index of the strong gauge group, and the subscript $L$
indicates left-handed helicities.

If I ignore the strong and electromagnetic interactions, leaving only
the weak $SU(2)$, each of these four doublets is equivalent and
independent.  I now arbitrarily pick two of them and do a charge
conjugation operation, thus working with their anti-particles
\begin{equation}\matrix{
\pmatrix{{u^g} \cr {d^g}\cr}_L \longrightarrow 
\pmatrix{\overline{{d^g}} \cr \overline{{u^g}}\cr}_R \cr
\pmatrix{{u^b} \cr {d^b}\cr}_L \longrightarrow 
\pmatrix{\overline{{d^b}} \cr \overline{{u^b}}\cr}_R \cr
}
\end{equation}
In four dimensions anti-fermions have the opposite helicity; so, I
label these new doublets with $R$ representing right handedness.

With two left and two right handed doublets, I combine them into Dirac
doublets
\begin{equation}
\pmatrix{
\pmatrix{\nu \cr e^-\cr}_L\cr
\pmatrix{\overline{{d^g}} \cr \overline{{u^g}}\cr}_R\cr
}
\qquad
\pmatrix{
\pmatrix{{u^r} \cr {d^r}\cr}_L\cr
\pmatrix{\overline{{d^b}} \cr \overline{{u^b}}\cr}_R \cr
}
\end{equation}
Formally in terms of the underlying fields, the construction takes
\begin{equation}\matrix{
\psi=\half (1-\gamma_5)\psi_{(\nu,e^-)}+\half (1+\gamma_5)
\psi_{({\overline{d^g}},{\overline{u^g}})} \cr
\chi=\half (1-\gamma_5)\psi_{({u^r}, {d^r})}+\half (1+\gamma_5)
\psi_{({\overline{d^b}},{\overline{u^b}})} \cr
}
\end{equation}

From the conventional point of view these fields have rather peculiar
quantum numbers.  For example, the left and right parts have different
electric charges.  Electromagnetism now violates parity.  The left and
right parts also have different strong quantum numbers; the strong
interactions violate parity as well.  Finally, the components have
different masses; parity is violated in the Higgs mechanism.  Making
the $SU(2)$ vector-like forces parity violation into all the other
interactions.

The different helicities of these fields also have variant baryon
number.  This is directly related to the known baryon violating
processes through weak ``instantons'' and axial
anomalies.\cite{thooft}  As discussed earlier, the axial anomaly
arises from a level flow out of the Dirac sea.\cite {anomalyflow}
This generates a spin flip in the fields, {\it i.e.} $e^-_L
\rightarrow ({\overline{u^g}})_R$.  Because of my peculiar particle
identification, this does not conserve charge, with $\Delta Q=
-{2\over 3} +1={1\over 3}$.  This would be a disaster for
electromagnetism were it not for the other Dirac doublet
simultaneously flipping, {\it i.e.} {${d^r}_L \rightarrow
({\overline{u^b}})_R$}, with a compensating $\Delta Q = -{1\over 3}$.
This is anomaly cancelation, with the total $\Delta Q = {1\over
3}-{1\over 3}=0$.  Only when both doublets are considered together is
the $U(1)$ symmetry restored.  In the overall process baryon number
remains violated, with $L+Q\rightarrow \overline Q + \overline Q$.
This is the famous `` `t Hooft vertex.''\cite {thooft}

This discussion has been in the continuum.  Now I return to the
lattice, and use the Kaplan-Shamir approach for fermions.\cite{kaplan}
\cite{shamir} \cite{mcih}  In this picture, our four dimensional
world is a ``4-brane'' embedded in 5-dimensions.  The complete lattice
is a five dimensional box with open boundaries, and the parameters are
chosen so the physical quarks and leptons appear as surface zero
modes.

\begin{figure}
\epsfxsize .6 \hsize
\centerline{\epsffile{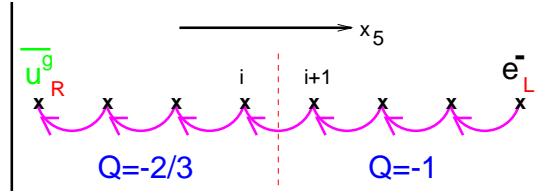}}
\caption{Pairing the electron with the anti-green-up-quark.}
\label{fig:1}
\end{figure}

I now insert the above pairing into this five dimensional scheme.  In
particular, I consider the left handed electron as a zero mode on one
wall and the right-handed anti-green-up quark as the partner mode on
the other wall, as sketched in Fig.~\ref{fig:1}.  This provides a
lattice regularization for the $SU(2)$ of the weak interactions.

However, since these two particles have different electric charge,
$U(1)_{EM}$ must be broken in the interior of the extra dimension.  I
now proceed in analogy to the ``waveguide'' picture\cite{waveguide}
and restrict this charge violation to $\Delta Q$ to one layer at some
interior $x_5=i$.  Then the fermion hopping term from $x_5=i$ to $i+1$
\begin{equation}
\overline\psi_{i}P\psi_{i+1}\qquad{(P=\gamma_5+r)}
\end{equation}
is a $Q=1/3$ operator.  At this layer, electric charge is not
conserved.  This is unacceptable and needs to be fixed.

\begin{figure}
\epsfxsize .6 \hsize
\centerline{\epsffile{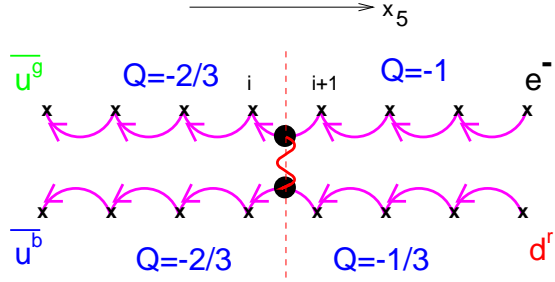}}
\caption {Transferring charge between the doublets introduces a
four-fermion coupling.  }
\label{fig:2}
\end{figure}

To restore the $U(1)$ symmetry I must transfer the charge from $\psi$ to
the compensating doublet $\chi$.  For this I replace the sum of
hoppings with a product on the offending layer
\begin{equation}
\overline\psi_{i}P\psi_{i+1}
{+}\overline\chi_{i}P\chi_{i+1}
\Black{\longrightarrow}
\overline\psi_{i}P\psi_{i+1}
{\times}\overline\chi_{i}P\chi_{i+1}
\end{equation}
This introduces an electrically neutral four fermi operator.  It is
explicitly baryon violating, involving a ``lepto-quark/diquark''
exchange, as sketched in Fig.~\ref{fig:2}.  One might think of the
operator as representing a ``filter'' at $x_5=i$ through which only
charge compensating pairs of fermions can pass.

In five dimensions there is no chiral symmetry.  Even for the free
theory, combinations like $\overline\psi_{i}P\psi_{i+1} $ have vacuum
expectation values.  I use such as a ``tadpole,'' with $\chi$
generating an effective hopping for $\psi$ and {\it vice versa}.

Actually the above four fermion operator is not quite sufficient for
all chiral anomalies, which can also involve right handed
singlet fermions.  To correct this I need explicitly include the right
handed sector, adding similar four fermion couplings
(also electrically neutral).

Having fixed the $U(1)$ of electromagnetism, I restore the strong
$SU(3)$ with an antisymmetrization $ {Q^r}{Q^g}{Q^b}{\longrightarrow
\epsilon^{\alpha\beta\gamma}Q^\alpha Q^\beta Q^\gamma}$.  Although the
quarks reside at different locations in the fifth dimension, this is
irrelevant since the $SU(3)$ symmetry need only be local in
four-dimensional space-time.  As for the electromagnetic case,
additional left-right inter-sector couplings are needed to correctly
obtain the effects of topologically non-trivial strong gauge fields.
These are of the same form as the strong 't Hooft vertex.

An alternative view folds the lattice about the interior of the fifth
dimension, placing all light modes on one wall and having the
multi-fermion operator on the other.  This is the model of Ref.~\cite
{us}, with the additional inter-sector couplings correcting a
technical error.\cite{neuberger}

Unfortunately the scheme is still non rigorous.  The most serious
worry is that the four fermion coupling might induce an unwanted
spontaneous symmetry breaking of one of the gauge symmetries.  I need
a strongly coupled paramagnetic phase without spontaneous symmetry
breaking.\cite{yukawa}  Ref.~\cite{us} showed how strongly coupled
zero modes preserve the desired symmetries, but the analysis ignored
contributions from heavy modes in the fifth dimension.

Assuming all works as desired, the model raises several interesting
questions.  A variation using a Majorana mass term on the extra wall
seems quite promising for formulating supersymmetric Yang-Mills theory
on the lattice.\cite{susy} Can a related scheme give a natural
formulation for more general supersymmetric theories?  Above I needed
a right handed neutrino to provide all quarks with partners.  Is there
some variation to avoids this particle, which completely decouples in
the continuum limit?  Another question concerns numerical simulations;
is the effective action positive?  Finally, I have used details of the
usual standard model, leaving open the question of whether this model
is somehow special.  Can I always use multi-fermion couplings to
eliminate undesired modes in other anomaly free chiral theories?
There is much more to do!

\section*{Acknowledgments}
This paper is based on a talk given at the February 1998 ``Workshop on
Nonperturbative Methods in Quantum Field Theory,'' sponsored by the
Australian National Institute for Theoretical Physics and the Special
Research Center for the Subatomic Structure of Matter at the
University of Adelaide.  This manuscript has been authored under
contract number DE-AC02-98CH10886 with the U.S.~Department of Energy.
Accordingly, the U.S.~Government retains a non-exclusive, royalty-free
license to publish or reproduce the published form of this
contribution, or allow others to do so, for U.S.~Government purposes.

\end{document}